\documentclass[submission,copyright,creativecommons]{eptcs}
\providecommand{\event}{ICLP 2024} % Name of the event you are submitting to

\usepackage{iftex}

\ifpdf
  \usepackage{underscore}         % Only needed if you use pdflatex.
  \usepackage[T1]{fontenc}        % Recommended with pdflatex
\else
  \usepackage{breakurl}           % Not needed if you use pdflatex only.
\fi

\usepackage{amsmath}
\usepackage{graphicx}
\usepackage{multirow}
\usepackage{amsthm}
\usepackage{amssymb}
\usepackage{tabularx}
\usepackage[ruled,noend,linesnumbered]{algorithm2e}
\usepackage{subcaption}
\usepackage{tikz}
\usetikzlibrary{shapes.geometric, arrows.meta, positioning}

\newtheorem{example}{Example}[section]
\usepackage{xcolor}

\title{Abduction of Domain Relationships from Data for VQA}
\author{Al Mehdi Saadat Chowdhury, Paulo Shakarian
\institute{School of Computing and Augmented Intelligence\\Arizona State University\\Tempe, Arizona, USA}
\email{\{achowd43,pshak02\}@asu.edu}
\and
Gerardo I. Simari
\institute{DCIC Univ.\ Nac.\ del Sur (UNS) \\ 
ICIC (UNS-CONICET) \\Bahia Blanca, Argentina}
\email{gis@cs.uns.edu.ar}
}
\def\titlerunning{Abduction of Domain Relationships from Data for VQA}
\def\authorrunning{A. M. S. Chowdhury, P. Shakarian, \& G. I. Simari}

\begin{document}
\maketitle

\begin{abstract}
In this paper, we study the problem of visual question answering (VQA) where the image and query are represented by ASP programs that lack domain data.  We provide an approach that is orthogonal and complementary to existing knowledge augmentation techniques where we abduce domain relationships of image constructs from past examples.
After framing the abduction problem, we provide a baseline approach, and an implementation that significantly improves the accuracy of query answering yet requires few examples.
\end{abstract}

\section{Introduction}
Visual Question Answering (VQA) is an AI task designed to reason about images.  Commonly, the image is transformed into a ``scene graph'' that enables the deployment of more formal reasoning tools.  For example, in recent work, both the scene graph and associated query were represented as an ASP Program~\cite{eiter2022neuro, kinjal2020aqua}; however,  notably the scene graph itself only contains information about the scene, but lacks commonsense knowledge -- in particular, knowledge about the domains of attributes identified by the scene. 
Existing work to address this shortcoming relies on leveraging large commonsense knowledge graphs for obtaining domain knowledge~\cite{marino2019ok, schwenk2022okvqa, wang2017fvqa}.  However, such approaches require the ability to accurately align the language of the knowledge graph with the language of the scene graph.  Further, for some applications, this does not guarantee that the aligned knowledge graph will necessarily improve VQA performance (e.g., if domain knowledge relevant to the queries is not possessed in the knowledge graph).  In this paper, we provide an orthogonal and complementary approach that leverages logical representations of the scene graph and query to abduce domain relationships that can improve query answering performance.  We frame the abduction problem and provide a simple algorithm that provides a valid solution. 
We also provide an implementation and show on a standard dataset that we can improve question answering accuracy from $59.98\%$ to $81.01\%$, and provide comparable results with few historical examples.

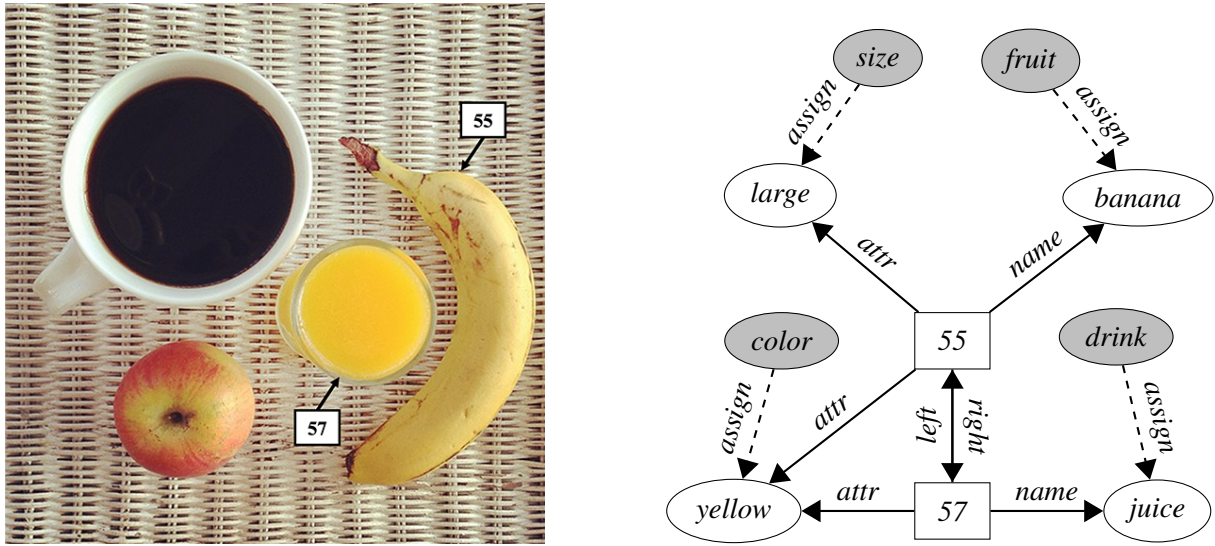
\begin{figure}[t]
    \centering
    \begin{subfigure}[t]{0.45\textwidth}
        \centering
        \includegraphics[width=\textwidth]{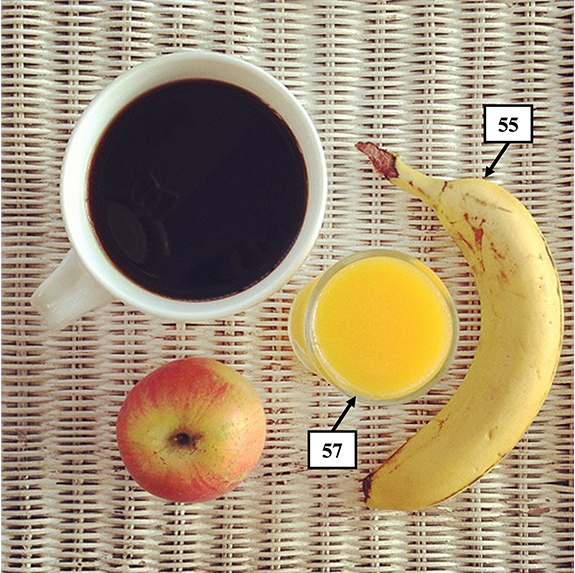}
        \label{fig:im}
    \end{subfigure}
    \hfill
    \begin{subfigure}[t]{0.45\textwidth}
        \centering
        \begin{tikzpicture}[node distance=1cm, auto, font=\itshape]
        % Initial Styles 
        \tikzstyle{rect} = [rectangle, draw, minimum width=1cm, minimum height=0.75cm]
        \tikzstyle{oval} = [ellipse, draw, minimum width=1cm, minimum height=0.75cm]
        \tikzstyle{dom} = [oval, fill=gray!50, text=black]
        \tikzstyle{arrow} = [-{Triangle[scale=1.5]}, thick]
        \tikzstyle{dashedarrow} = [arrow, dashed]
        
        % Objects 
        \node[rect] (n57) {57};
        \node[rect, above=1.5cm of n57] (n55) {55};
        
        % Attributes for n57
        \node[oval, right=1.25cm and 1.5cm of n57] (juice) {juice};
        \node[oval, left=1.5cm and 1.5cm of n57] (yellow) {yellow};
        \node[dom, left=0.5cm and 1.0cm of n55] (color) {color};
        \node[dom, above left=1.75cm and -0.5cm of juice] (drink) {drink};
        
        % Attributes for n55
        \node[oval, above left=1.25cm and 1.25cm of n55] (large) {large};
        \node[oval, above right=1.25cm and 1.25cm of n55] (banana) {banana};
        \node[dom, above left=1.25cm and -2.25cm of large] (size) {size};
        \node[dom, above left=1.25cm and 0.25cm of banana] (fruit) {fruit};
    
        % Edges from n57 to its children
        \draw[arrow] (n57) -- (yellow) node[midway, above, sloped] {attr};
        \draw[arrow] (n57) -- (juice) node[midway, above, sloped] {name};
        \draw[dashedarrow] (color) -- (yellow) node[midway, above, sloped] {assign};
        \draw[dashedarrow] (drink) -- (juice) node[midway, above, sloped] {assign};
        
        % Edges from n55 to its children
        \draw[arrow] (n55) -- (yellow) node[midway, above, sloped] {attr};
        \draw[arrow] (n55) -- (banana) node[midway, above, sloped] {name};
        \draw[arrow] (n55) -- (large) node[midway, above, sloped] {attr};
        \draw[dashedarrow] (fruit) -- (banana) node[midway, above, sloped] {assign};
        \draw[dashedarrow] (size) -- (large) node[midway, above, sloped] {assign};
    
        % Relationships
        \draw[arrow] (n55) -- (n57) node[midway, above, sloped] {right};
        \draw[arrow] (n57) -- (n55) node[midway, above, sloped] {left};
        \end{tikzpicture}
        \label{fig:sg}
    \end{subfigure}
    \caption{An image (left) and a section of its corresponding scene graph (right). In the scene graph, square nodes represent objects, oval nodes represent attributes, and solid edges connect objects to attributes. Shaded nodes represent domain knowledge, connected to attributes by dashed edges.}
    \label{fig:sgaug}
\end{figure}

\medskip
\noindent\textbf{Motivating Example.} Consider the simple scene graph depicted in Figure \ref{fig:sgaug} and the query ``\textit{What is the color of the fruit to the right of the juice?}''.
Without the shaded nodes (which indicate domain information external to the image) there is no attribute of any constant associated with \textit{banana} that is associated with the domain \textit{color} or the domain \textit{fruit}.  Hence, the only answer would be to assume that there is no fruit or the color information is not given, or randomly guess \textit{large} 
(while not a color, it is an attribute) or \textit{yellow}.  
In this paper, we will look to abduce these domain relationships from a limited number of examples.

\section{Technical Preliminaries}
We extend the framework of~\cite{eiter2022neuro}, which represents both images and queries as ASP programs (and the programs can be directly represented as an equivalent scene graph as shown in Figure~\ref{fig:sgaug}).  Their approach to VQA leverages a neurosymbolic framework and was tested on synthetic datasets (e.g., CLEVR~\cite{johnson2017clevr}) that involve limited objects and attributes.
We seek to extend their results to real-world datasets such as GQA~\cite{hudson2019gqa}, which are more complex. We follow the logic programming construct as~\cite{eiter2022neuro} in that we have logical facts representing the scene graphs ($\Pi^{I}$), the query to be answered ($\Pi^{Q}$), as well as standard ``VQA helper'' rules ($\Pi^{R}$).

We assume the existence of a first order logical language (constants $\mathcal{C}$, variables $\mathcal{V}$, predicates $\mathcal{P}$).  
Set $\mathcal{C}$ has several subsets: objects ($\mathcal{C}_{obj}$), attributes ($\mathcal{C}_{att}$), domains ($\mathcal{C}_{dom}$), and single choice questions ($\mathcal{C}_{sinChoice}$).  Additionally, we will have a special binary predicate $assign$ where the first argument is an attribute and the second is a domain.  
Every attribute can thus be associated with one or more domains via atom $assign(a,d)$, meaning that attribute $a$ has domain $d$. We will also define Answer Set Programming (ASP) rules in the usual manner; a rule with no body is a fact and a set of rules is a program.  Given a program $\Pi$, the subset of facts in $\Pi$ where the head is formed with $assign$ is called the ``domain relationships'', and denoted $\Pi^D$.  
Likewise, we assume programs representing an image and a query, $\Pi^I$ and $\Pi^Q$, respectively, that do not contain domain relationships, and a common set of rules $\Pi^R$ that answers the query using $\Pi^I$ and $\Pi^Q$.  
Also, we shall use the standard ASP semantics based on interpretations~\cite{eiter2022neuro}, and use the notation $I \models \Pi$ to denote that interpretation $I$ satisfies program $\Pi$.  
Further, we say that program $\Pi_1 \models \Pi_2$ (read ``$\Pi_1$ entails $\Pi_2$'') meaning that all interpretations that satisfy $\Pi_1$ also satisfy $\Pi_2$.

In this work, we are primarily concerned with the case where there is a common $\Pi^D$ for a collection of image-query program pairs (``examples'') denoted 
$\langle\Pi^{I}_1, \Pi^Q_1\rangle, \ldots, \langle\Pi^{I}_n, \Pi^Q_n\rangle$.  
We may also know that a given $\langle\Pi^{I}_i,\Pi^Q_i\rangle$ is associated with some set of {\em ground truth} $\Pi^{GT}_i$. Due to the lack of domain knowledge, $\Pi^{I}_i\cup\Pi^Q_i\cup\Pi^{R}$ may not entail $\Pi^{GT}_i$. 
However if an oracle provides a correct $\Pi^D$, we have that $\Pi^{I}_i\cup\Pi^{Q}_i \cup \Pi^{R} \cup \Pi^{D} \models \Pi^{GT}_i$. 
We show an example of this case below taken from the scene graph dataset of~\cite{hudson2019gqa} (depicted in Figure~\ref{fig:sgaug}), which we also use in our experiments.

\begin{example}
    Consider a program $\Pi_i = \Pi^{I}_{i} \cup \Pi^{Q}_{i} \cup \Pi^{R}$ that consists of the following scene representation~$\Pi^{I}_{i}$, question representation $\Pi^{Q}_{i}$ for the question ``What is the color of the fruit to the right of the juice?'', and the set of rules $\Pi^{R}$ common to all image-query program pairs:

    \begin{equation*}
        \begin{array}{cc}
            \Pi^{I}_{i} = \left\{
            \begin{array}{lllll}
                ob(2317538, 51). &
                name(51, cup). &
                attr(51, glass). & 
                attr(51, white). \\
                ob(2317538, 54).  &
                name(54, apple).  &
                attr(54, round).  &
                attr(54, red).  \\
                ob(2317538, 55).  &
                name(55, banana).  &
                attr(55, yellow).  & attr(55, large). & rel(55,57,right). \\ 
                ob(2317538, 57).  &
                name(57, juice).  &
                attr(57, yellow).  & rel(57,55,left). \\ 
            \end{array}
            \right.
            \end{array}
    \end{equation*}
    
    \begin{equation*}
        \begin{array}{cc}
            \Pi^{Q}_{i} = \left\{
            \begin{array}{lll}
                scene(0, 2317538).  &
                select(1,juice,0).  &
                relate(2,fruit,right,1).  \\
                query(4,color,3).  &
                exit(5). 
            \end{array}
            \right.
        \end{array}
    \end{equation*}

\noindent
As in~\cite{eiter2022neuro}, our question representation $\Pi^{Q}_i$ is structured so that each query part is organized sequentially, with the first argument of each predicate indicating order and the last argument showing dependency on prior results. This step-by-step approach along with $\Pi^{R}$ aids in answering questions effectively:

    \begin{equation*}
        \begin{array}{c}
            \Pi^{R} = \left\{
            \begin{array}{rcl}
                r(T, OID) &:-& scene(T,S), ob(S,OID). \\
                r(T, OID) &:-& select(T, ON, D), r(D, OID), name(OID, ON). \\
                r(T, TID) &:-& relate(T, GC, R, D), r(D, OID), rel(TID, OID, R), name(TID, ON), \\& & assign(ON,GC). \\
                r(T, A) &:-& query(T, color, D), r(D, OID), attr(OID, A), assign(A, color). \\
                result(RSLT) &:-& exit(T), r(T-1, RSLT). \\
                empty(AT) &:-& exit(T), not\ r(AT,\_), AT=0..T-1.
            \end{array}
            \right.
        \end{array}
    \end{equation*}

    \noindent
    For this question, the ground truth is the program: 

    \begin{equation*}
        \begin{array}{c}
            \Pi^{GT}_{i} = \left\{
            \begin{array}{rcl}
                result(yellow).
            \end{array}
            \right\}
        \end{array}
    \end{equation*}

    \noindent
    However, due to the lack of atoms $assign(banana, fruit)$ and $assign(yellow, color)$, we see that, 
    \mbox{$\Pi_{i} \nvDash \Pi^{GT}_{i}$}. Now we assume that an oracle provides us with $\Pi^{D}$, as follows:
    \begin{equation*}
        \begin{array}{cc}
             \Pi^{D} = \left\{
            \begin{array}{lll}
                assign(glass, material). &
                assign(white, color). &
                assign(apple, fruit). \\
                assign(round, shape). &
                assign(red, color). &
                assign(banana, fruit). \\
                assign(yellow, color). & assign(large,size). &
                assign(juice, drink).
            \end{array}
            \right.
        \end{array}
    \end{equation*}

\noindent
    With the existence of this domain $\Pi^{D}$, now we have $\Pi_{i} \cup \Pi^D \models \Pi^{GT}_{i}$.
\label{ex:techprelim}
\end{example}

\smallskip
\noindent\textbf{Fallback Rules.}
In this framework, where we may have an absent or partial $\Pi^D$, it is useful to have ``fallback rules'' of the form: 
$assign(att,default) \leftarrow \bigwedge_{att \in \mathcal{C}_{att}\setminus\{default\}}\neg assign(att,DOM)$.  This assumes a special attribute constant ``default'' to which an object without an attribute falls back.  
The next example augments Example~\ref{ex:techprelim} with fallback rules:

\smallskip

\begin{example}
    We assume additional fallback rules, added to $\Pi^{R}$, of the form:
        \begin{equation*}
        \begin{array}{c}
           % \Pi^{FB} = \left\{
            \begin{array}{rcl}
                r(T, A) &:-& query(T, color, D), r(D, OID), attr(OID, A),
                \\ && \neg assign(A, color), assign(A, default).
            \end{array}
            %\right.
        \end{array}
    \end{equation*}
    Returning to our running example, assuming there is no $\{assign(yellow,color).\} \in \Pi^{D}$, adding fallback rules, we get the following $\Pi^{D}$:
    \begin{equation*}
        \begin{array}{cc}
             \Pi^{D} = \left\{
            \begin{array}{lll}
                assign(glass, material). &
                assign(white, color). &
                assign(apple, fruit). \\
                assign(round, shape). &
                assign(red, color). &
                assign(banana, fruit). \\
                assign(yellow, default). & assign(large,size). &
                assign(juice, drink).
            \end{array}
            \right.
        \end{array}
    \end{equation*}
\label{ex:prwithfallback}
\end{example}

\noindent\textbf{Abducing Domain Relationships.}  We now formalize our problem.  
Given examples
$\mathbf{EX} = \{\langle \Pi^{I}_1,\Pi^Q_1\rangle,$ $\ldots,\langle\Pi^{I}_n,\Pi^Q_n\rangle \}$ with a common rule set $\Pi^{R}$ (which may or may not include fallback rules) and corresponding ground truth $\mathbf{GT}=\{\Pi^{GT}_1,\ldots,\Pi^{GT}_n \}$, 
then $\langle \mathbf{EX},\mathbf{GT},\Pi^{R} \rangle$ is a \textit{domain abduction problem} (DAP).  

Any $\Pi^D$ containing only facts formed with $assign$ in the head is a {\em hypothesis} for a DAP.  
A hypothesis $\Pi^D$ is an \textit{explanation} for DAP $\langle \mathbf{EX},\mathbf{GT},\Pi^{R} \rangle$ if and only if for all $i$ we have $\Pi^{I}_i\cup\Pi^{Q}_i\cup\Pi^{R}\cup\Pi^D \models  \Pi^{GT}_i$.  
However, when $\mathbf{EX},\mathbf{GT}$ are noisy (e.g., produced from a machine learning system) there may be no explanation; in such cases, we may be able to find a hypothesis $\Pi^D$ that maximizes some accuracy or recall metric.
For example, finding $\Pi^D$ that maximizes $\frac{1}{|\mathbf{GT}|} |\{\Pi^{GT}_i \in \mathbf{GT} \textit{ s.t. } \Pi^{I}_i\cup\Pi^{Q}_i\cup\Pi^{R}\cup\Pi^D \models  \Pi^{GT}_i\}|$ (where $|\cdot|$ is set cardinality) would lead to maximized accuracy.

\section{A Practical Heuristic Algorithm}
\begin{algorithm}[t]
    % \LinesNumbered
    % \SetAlgoLined
    \DontPrintSemicolon
    
    \SetKwInOut{Input}{Input}
    \SetKwInOut{Output}{Output}
    
    \SetKwFunction{Prune}{PRUNE}

    \Input{A set of programs $\mathbf{EX}=\{\langle \Pi^{I}_1,\Pi^Q_1\rangle,\ldots,\langle\Pi^{I}_n,\Pi^Q_n\rangle \}$ where $\Pi^{I}_i$, and $\Pi^{Q}_i$ correspond to scene and question representation;
    Common Rule Set $\Pi^{R}$ with Fallback rules; \\
    Set of ground truths $\mathbf{GT} = \{\Pi^{GT}_1,\ldots,\Pi^{GT}_n\}$.}
    \Output{A hypothesis $\Pi^{D}$}

    $\Pi^{D} \leftarrow \emptyset$ \;
    \ForEach{$\langle \Pi^{I}_i,\Pi^Q_i\rangle \in \mathbf{EX}$}{
        \If(\tcp*[h]{$w$ is the query type, $x, y$ are possible answers}){$choose(w,x,y) \in \Pi^{Q}_i$}{ \label{a:choosestart}
            \If{$w \in \mathcal{C}_{sinChoice}$}{
                \lIf{$result(x) \in \Pi^{GT}_i$}{$\Pi^{D} \leftarrow \Pi^{D} \cup \{assign(x,w).\}$}
                \lElse{$\Pi^{D} \leftarrow \Pi^{D} \cup \{assign(y,w).\}$}
            }
            \Else{
                $\Pi^{D} \leftarrow \Pi^{D} \cup \{assign(x,w).\ assign(y,w).\}$
            }
        } \label{a:chooseend}
        \If{$\Pi^{I}_i\cup\Pi^Q_i\cup\Pi^{R} \nvDash \Pi^{GT}_i$}{ \label{a:gtsbeg}
            Pick the fact $select(i,c,j) \in \Pi^{Q}_i$ such that $\Pi^{I}_i\cup\Pi^Q_i\cup\Pi^{R} \models empty(i)$ and $i$ is minimal \;
            \If(\tcp*[h]{$c$ is then a general concept}){there does not exist $name(\_,c) \in \Pi^{I}_i$}{
                Pick $c' \neq c$ such that $name(\_,c') \in \Pi^{I}_i$ and $\Pi^{I}_i\cup\Pi^Q_i\cup\Pi^{R} \cup \{assign(c',c).\} \models \Pi^{GT}_i$ \;
                $\Pi^{D} \leftarrow \Pi^{D} \cup \{assign(c',c).\}$ \;
                $support_{c',c}$ += $1$ \;
            }
        }
    }  \label{a:gtsend}
    \Return $\{assign(c',c). \in \Pi^{D}$ with $support_{c',c} > threshold$\} \;
    \caption{FAST-DAP}
    \label{algo:train}
\end{algorithm}

In this section, we present a practical, heuristic algorithm for finding a DAP, that while is not guaranteed to maximize the accuracy of question answering, we show to perform very well in practice.  There are several reasons as to why we adopt this more practical approach.  First, in the general case, a brute-force approach is intractable.  Second, even if it is possible to exactly optimize an accuracy metric as described in the previous section, it may still perform poorly when confronted with unseen data due to overfitting.  Third, in some cases, the query itself can reveal portions of the ground truth.  To address all of these issues, we introduce our practical heuristic algorithm FAST DAP (Algorithm~\ref{algo:train}).
Regarding the first point, the algorithm is highly performant, requiring only one pass over all examples in $\textbf{EX}$ -- this also allows for trivial parallelization.  Second, we only add facts to $\Pi^D$ that support a certain number of examples, which acts as a form of regularization; we then tune this threshold to maximize accuracy.  
To address the third point, in lines~\ref{a:choosestart}--\ref{a:chooseend} we utilize examples that provide domain information in the query itself (with two answers as in $choose(color, red, blue, 0)$ and with single answer as in $choose(healthy, apple, cake, 0$)), while we leverage the step-by-step nature of the ASP formulation of queries (following~\cite{eiter2022neuro}, see Example~\ref{ex:techprelim}) to identify domain assignments that can satisfy the ground truth (lines \ref{a:gtsbeg}-\ref{a:gtsend}).

\section{Evaluation}
We now report on the results of our experimental evaluation. 
We use the GQA dataset~\cite{hudson2019gqa}, allowing us to build on the results of \cite{eiter2022neuro}, which uses the CLEVR~\cite{johnson2017clevr} synthetic data.  
Note that we use ground truth ASP representations of the images and queries.  
We examine our practical heuristic in four different ways.  
First, we examine the accuracy improvements when employing FAST-DAP. 
Second, we examine its data efficiency (e.g., how many examples in \textbf{EX} are required to provide useful results).  
Third, we examine the sensitivity of the support threshold for elements of $\Pi_D$. 
Finally, we examine running time.  
We created our implementation in Python~3.11.7 and use the Clingo solver for the ASP engine.
Experiments were run on an Apple~M2 machine with a~10-core CPU, and 32GB of RAM. 
All computations were carried out using only the CPU (the system’s GPU was not used).
We now present the results of each experiment.

\medskip
\noindent\textbf{Accuracy.}  We assess our approach's accuracy against the baseline (no $\Pi^D$), evaluating improvements with and without fallback rules (FBR and No FBR), both utilizing FAST-DAP. For the baseline (no FAST-DAP), the ASP solver either provides an answer or returns ``empty'' if it cannot deduce one. 
On our test set (disjoint from the examples), the baseline accuracy across all question types was~$59.98\%$ without domain information. Incorporating domain information learned from the training set significantly boosted accuracy to~$80.62\%$ without fallback rules, and $81.01\%$ with them.
To gain deeper insights, we analyze specific question types, a subset of which is presented in Table~\ref{tab:acc-indv}.
Some types, such as verification questions, show minimal dependence on domain categorization, while others rely more heavily on it. Additionally, certain questions require translating specific concepts into general terms (FAST-DAP, lines~\ref{a:gtsbeg}-\ref{a:gtsend}), like generalizing ``banana'' to ``fruit'' or ``juice'' to ``drink.'' In Table~\ref{tab:acc-indv}, all non-choice queries require such generalization.

\begin{table}[thb!]
    \setlength{\tabcolsep}{12pt}
    \centering
    \begin{tabular}{ m{3cm} m{2.5cm} m{2.5cm} m{2.5cm} }
        \hline
        \textbf{Question Type} & \textbf{Baseline} & \textbf{FBR (Ours)} & \textbf{No FBR (Ours)} \\
        \hline
        choose\_activity   & 69.02 & 95.11 & 94.84 \\
        choose\_color   & 89.80 & 93.48 & 93.21 \\
        choose\_older   & 0 & 97.24 & 97.24 \\
        choose\_rel   & 73.88 & 85.48 & 81.72 \\
        choose\_vposition   & 96.27 & 94.98 & 94.93 \\
        \hline
        and  & 94.25 & 91.93 & 91.83 \\
        verify\_age   & 86.89 & 97.54 & 97.54 \\
        verify\_color   & 95.71 & 96.58 & 96.44 \\
        verify\_location   & 49.28 & 94.5 & 94.5 \\
        query   & 36.07 & 72.83 & 72.20 \\
        \hline
    \end{tabular}
    \caption{Evaluation of answering questions. The ``Baseline'' column shows accuracy (in percentage) without learned domains, ``FBR'' shows accuracy with learned domains and fallback rules, and ``No FBR'' shows accuracy with domain atoms but without using fallback rules.}
    \label{tab:acc-indv}
\end{table}

\begin{figure}[t]
    \centering
    \begin{subfigure}[t]{0.45\textwidth}
        \centering
        \includegraphics[width=\textwidth]{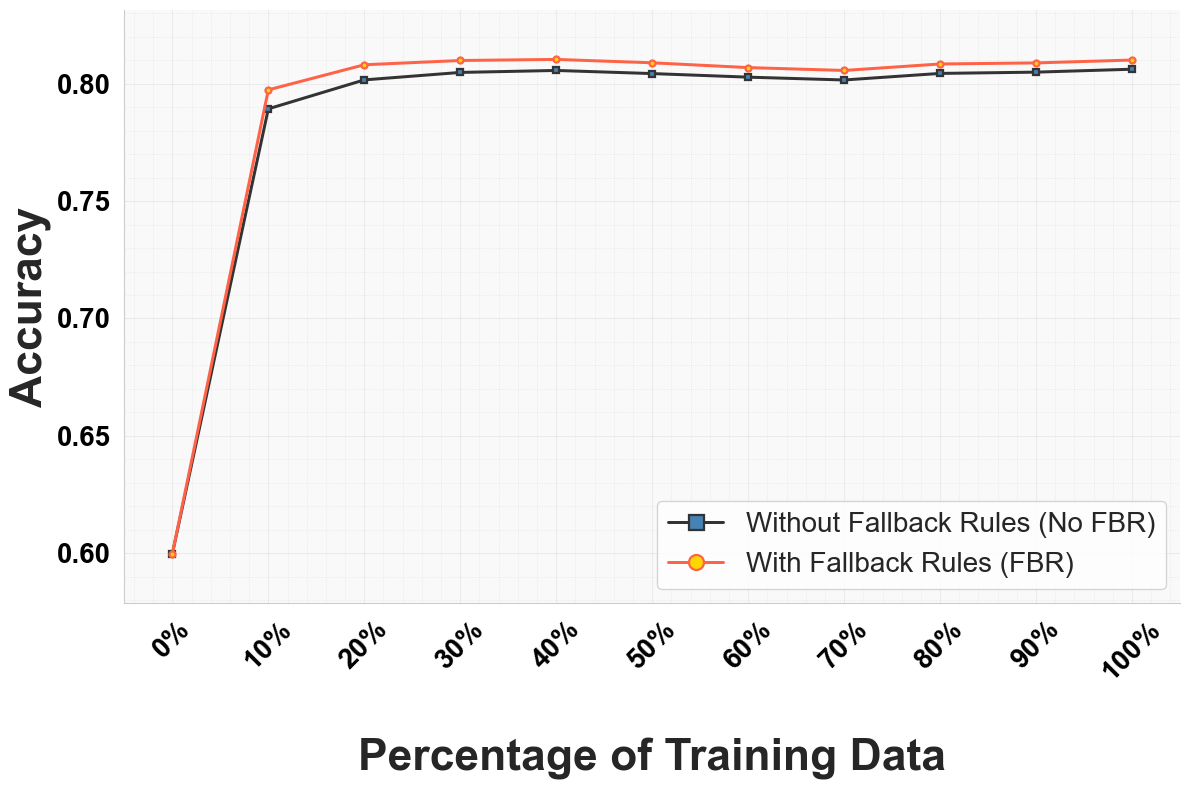}
        \caption{Accuracy on the test set leveraging learned domains from different training subsets. 
        }
        \label{fig:trainsubsets}
    \end{subfigure}
    \hfill
    \begin{subfigure}[t]{0.45\textwidth}
        \centering
        \includegraphics[width=\textwidth]{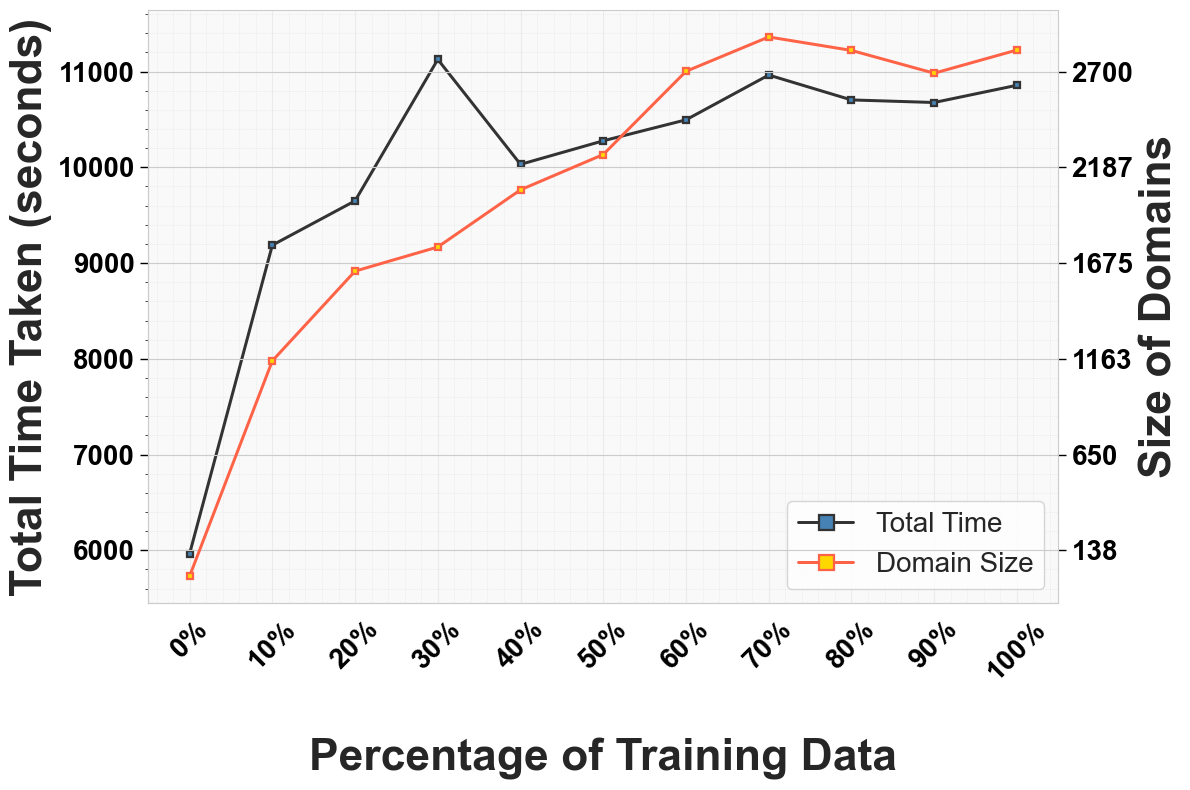}
        \caption{Execution time of our algorithm for different sample sizes, run in parallel with identical settings.
        }
        \label{fig:time}
    \end{subfigure}
    \caption{Accuracy and running time on different training subsets.}
    \label{fig:acctime}
\end{figure}

\medskip
\noindent\textbf{Data Efficiency.} In this second experiment, we aimed to find the optimal sample size for learning domains. We randomly divided the data as follows: 20\% for training, 10\% for validation, and the remaining 70\% for testing. Instead of using the entire training set at once, we divided it into~11 progressively larger subsets as follows:
the first subset served as a baseline model with no samples, the second subset contained 10\% of the training data, the third subset included the first 10\% plus an additional 10\%, making up 20\% of the training data, and this pattern continued until the 11th subset, which encompassed all the training data. 
Each training subset was used independently to learn the domains, and these learned domains were then used to predict the answers in the test set. Figure~\ref{fig:trainsubsets} illustrates the results, showing accuracy across the training data for two scenarios: the black line represents the learned domain without fallback rules, while the red line includes fallback rules. 
As depicted in Figure~\ref{fig:trainsubsets}, using just 10\% of the training set (equivalent to 2\% of the entire dataset) achieves a respectable accuracy of 78.93\%. 
With 20\% of the training data (4\% of the entire dataset), accuracy exceeds 80\%. 
This suggests that a small amount of data can effectively learn domains, with only slight accuracy gains from adding more data.

\medskip
\noindent\textbf{Threshold Sensitivity.}  FAST-DAP refines the learned domain set by removing domains whose support falls below a specified threshold. This approach helps regularize the outcome since the domains were derived from the application of possibly noisy data and rules. The threshold is a hyper-parameter determined from the validation set. 
We used the $10^{th}$ to $70^{th}$ percentile support values as potential thresholds. For each, we removed domains with lower support, assessed validation accuracy, and selected the threshold with the highest accuracy. Domains below this final threshold were then removed.
Table \ref{tab:hs} illustrates the accuracy achieved at different thresholds. Based on this data, we selected a threshold of $59.5$, and domains with support below this value were excluded to form the final set of domains.

\begin{table}[t]
    \begin{subtable}[t]{0.45\textwidth}
        \centering
        \begin{tabular}{ m{1.5cm} m{1.5cm} m{1.5cm} }
            \hline
            \textbf{Percentile} & \textbf{Threshold} & \textbf{Accuracy} \\
            \hline
            10   & 12.3 & 79.44 \\
            20   & 20.6 & 79.79\\
            30   & 30.9 & 79.89\\
            40   & 46.4 & 80.10\\
            \hline
        \end{tabular}
    \end{subtable}
    \begin{subtable}[t]{0.45\textwidth}
        \centering
        \begin{tabular}{ m{1.5cm} m{1.5cm} m{1.5cm} }
            \hline
            \textbf{Percentile} & \textbf{Threshold} & \textbf{Accuracy} \\
            \hline
            50   & \textbf{59.5} & \textbf{80.54}\\
            60   & 90.8 & 80.02\\
            70   & 121.2 & 79.75\\
                 &      &      \\
            \hline
        \end{tabular}
    \end{subtable}
    \caption{Accuracy results on the validation set after removing domains with support below a threshold.
    }
    \label{tab:hs}
\end{table}

\medskip
\noindent\textbf{Running Time.}  
The running time of our algorithm is primarily influenced by the performance of the ASP solver Clingo, and is directly proportional to the number of atoms it processes. 
Figure~\ref{fig:time} illustrates that the running time grows consistently from the base case with no training samples to the scenario where all training samples are used. 
Incorporating more training samples to learn domains substantially boosts the number of learned new domain atoms, thereby requiring Clingo to process more atoms during deduction. This necessity is the main factor driving the increase in running time. However, note that this increase is bounded by a constant factor related to the domain's size.

\section{Conclusion}
In this paper, we introduced a practical heuristic algorithm designed to infer domain relationships from a logical representation of data specifically for visual question answering. Our algorithm is highly efficient, requiring just a single pass over the data, and it significantly enhances accuracy compared to using a logical representation that does not leverage domain information.
Despite its strong practical performance, an important limitation of our approach is that there are no theoretical guarantees for the solutions it obtains.
A promising direction for future research focused on addressing this limitation is to refine our approach by incorporating meta-cognitive AI~\cite{wei2024met} techniques.

\section*{Acknowledgement}
This research was funded by Army Research Office (ARO) grant W911NF-24-1-0007.

\bibliographystyle{eptcs}
\bibliography{generic}

\begin{thebibliography}{1}
\providecommand{\bibitemdeclare}[2]{}
\providecommand{\surnamestart}{}
\providecommand{\surnameend}{}
\providecommand{\urlprefix}{Available at }
\providecommand{\url}[1]{\texttt{#1}}
\providecommand{\href}[2]{\texttt{#2}}
\providecommand{\urlalt}[2]{\href{#1}{#2}}
\providecommand{\doi}[1]{doi:\urlalt{https://doi.org/#1}{#1}}
\providecommand{\eprint}[1]{arXiv:\urlalt{https://arxiv.org/abs/#1}{#1}}
\providecommand{\bibinfo}[2]{#2}

\bibitemdeclare{inproceedings}{kinjal2020aqua}
\bibitem{kinjal2020aqua}
\bibinfo{author}{Kinjal \surnamestart Basu\surnameend}, \bibinfo{author}{Farhad
  \surnamestart Shakerin\surnameend} \& \bibinfo{author}{Gopal \surnamestart
  Gupta\surnameend} (\bibinfo{year}{2020}): \emph{\bibinfo{title}{AQuA:
  ASP-Based Visual Question Answering}}.
\newblock In: {\slshape \bibinfo{booktitle}{PADL}},
  \bibinfo{publisher}{Springer International Publishing},
  \bibinfo{address}{Cham}, pp. \bibinfo{pages}{57--72},
  \doi{10.1007/978-3-030-39197-3_4}.

\bibitemdeclare{article}{eiter2022neuro}
\bibitem{eiter2022neuro}
\bibinfo{author}{Thomas \surnamestart Eiter\surnameend},
  \bibinfo{author}{Nelson \surnamestart Higuera\surnameend},
  \bibinfo{author}{Johannes \surnamestart Oetsch\surnameend} \&
  \bibinfo{author}{Michael \surnamestart Pritz\surnameend}
  (\bibinfo{year}{2022}): \emph{\bibinfo{title}{A neuro-symbolic ASP pipeline
  for visual question answering}}.
\newblock {\slshape \bibinfo{journal}{TPLP}}
  \bibinfo{volume}{22}(\bibinfo{number}{5}), pp. \bibinfo{pages}{739--754},
  \doi{10.1017/S1471068422000229}.

\bibitemdeclare{inproceedings}{hudson2019gqa}
\bibitem{hudson2019gqa}
\bibinfo{author}{Drew~A. \surnamestart Hudson\surnameend} \&
  \bibinfo{author}{Christopher~D. \surnamestart Manning\surnameend}
  (\bibinfo{year}{2019}): \emph{\bibinfo{title}{GQA: A New Dataset for
  Real-World Visual Reasoning and Compositional Question Answering}}.
\newblock In: {\slshape \bibinfo{booktitle}{CVPR}}, pp.
  \bibinfo{pages}{6693--6702}, \doi{10.1109/CVPR.2019.00686}.

\bibitemdeclare{inproceedings}{johnson2017clevr}
\bibitem{johnson2017clevr}
\bibinfo{author}{Justin \surnamestart Johnson\surnameend},
  \bibinfo{author}{Bharath \surnamestart Hariharan\surnameend},
  \bibinfo{author}{Laurens \surnamestart van~der Maaten\surnameend},
  \bibinfo{author}{Li~\surnamestart Fei-Fei\surnameend},
  \bibinfo{author}{C.~Lawrence \surnamestart Zitnick\surnameend} \&
  \bibinfo{author}{Ross \surnamestart Girshick\surnameend}
  (\bibinfo{year}{2017}): \emph{\bibinfo{title}{CLEVR: A Diagnostic Dataset for
  Compositional Language and Elementary Visual Reasoning}}.
\newblock In: {\slshape \bibinfo{booktitle}{CVPR}}, pp.
  \bibinfo{pages}{1988--1997}, \doi{10.1109/CVPR.2017.215}.

\bibitemdeclare{inproceedings}{marino2019ok}
\bibitem{marino2019ok}
\bibinfo{author}{Kenneth \surnamestart Marino\surnameend},
  \bibinfo{author}{Mohammad \surnamestart Rastegari\surnameend},
  \bibinfo{author}{Ali \surnamestart Farhadi\surnameend} \&
  \bibinfo{author}{Roozbeh \surnamestart Mottaghi\surnameend}
  (\bibinfo{year}{2019}): \emph{\bibinfo{title}{OK-VQA: A Visual Question
  Answering Benchmark Requiring External Knowledge}}.
\newblock In: {\slshape \bibinfo{booktitle}{CVPR}}, pp.
  \bibinfo{pages}{3190--3199}, \doi{10.1109/CVPR.2019.00331}.

\bibitemdeclare{inproceedings}{schwenk2022okvqa}
\bibitem{schwenk2022okvqa}
\bibinfo{author}{Dustin \surnamestart Schwenk\surnameend},
  \bibinfo{author}{Apoorv \surnamestart Khandelwal\surnameend},
  \bibinfo{author}{Christopher \surnamestart Clark\surnameend},
  \bibinfo{author}{Kenneth \surnamestart Marino\surnameend} \&
  \bibinfo{author}{Roozbeh \surnamestart Mottaghi\surnameend}
  (\bibinfo{year}{2022}): \emph{\bibinfo{title}{A-OKVQA: A Benchmark for Visual
  Question Answering Using World Knowledge}}.
\newblock In: {\slshape \bibinfo{booktitle}{ECCV}},
  \bibinfo{publisher}{Springer Nature Switzerland}, \bibinfo{address}{Cham},
  pp. \bibinfo{pages}{146--162}, \doi{10.1007/978-3-031-20074-8_9}.

\bibitemdeclare{article}{wang2017fvqa}
\bibitem{wang2017fvqa}
\bibinfo{author}{Peng \surnamestart Wang\surnameend},
  \bibinfo{author}{Qi~\surnamestart Wu\surnameend}, \bibinfo{author}{Chunhua
  \surnamestart Shen\surnameend}, \bibinfo{author}{Anthony \surnamestart
  Dick\surnameend} \& \bibinfo{author}{Anton \surnamestart van~den
  Hengel\surnameend} (\bibinfo{year}{2018}): \emph{\bibinfo{title}{FVQA:
  Fact-Based Visual Question Answering}}.
\newblock {\slshape \bibinfo{journal}{IEEE Transactions on Pattern Analysis and
  Machine Intelligence}} \bibinfo{volume}{40}(\bibinfo{number}{10}), pp.
  \bibinfo{pages}{2413--2427}, \doi{10.1109/TPAMI.2017.2754246}.

\bibitemdeclare{misc}{wei2024met}
\bibitem{wei2024met}
\bibinfo{author}{Hua \surnamestart Wei\surnameend}, \bibinfo{author}{Paulo
  \surnamestart Shakarian\surnameend}, \bibinfo{author}{Christian \surnamestart
  Lebiere\surnameend}, \bibinfo{author}{Bruce \surnamestart Draper\surnameend},
  \bibinfo{author}{Nikhil \surnamestart Krishnaswamy\surnameend} \&
  \bibinfo{author}{Sergei \surnamestart Nirenburg\surnameend}
  (\bibinfo{year}{2024}): \emph{\bibinfo{title}{Metacognitive AI: Framework and
  the Case for a Neurosymbolic Approach}}, \doi{10.48550/arXiv.2406.12147}.

\end{thebibliography}
\end{document}